\documentclass[a4paper,fleqn,usenatbib]{mnras}

\usepackage{newtxtext,newtxmath}
\usepackage[T1]{fontenc}
\usepackage{ae,aecompl}

\usepackage{graphicx}	% Including figure files
\usepackage{amsmath}	% Advanced maths commands
\usepackage{amssymb}	% Extra maths symbols
\usepackage{subfig}	% For figures position
\usepackage{epstopdf}	% To use .eps files with pdfLaTeX
\usepackage{multirow}	% To have multirows in columns

\newcommand*\dif{\mathop{}\!\mathrm{d}} 	% to have a proper differential symbol in integrals
\title[The mass dependence of dark matter halo alignments]{The mass dependence of dark matter halo alignments with large-scale structure}

\author[D. Piras et al.]{
Davide Piras$^{1, 2}$\thanks{E-mail: d.piras@ucl.ac.uk; davide.piras@studenti.unipd.it},
Benjamin Joachimi$^{1}$,
Bj\"{o}rn Malte Sch\"{a}fer$^{3}$,
Mario Bonamigo$^{4,5}$, 
\newauthor{Stefan Hilbert$^{6,7}$,
and Edo van Uitert$^{1}$}
\\
$^{1}$Department of Physics and Astronomy, University College London, Gower Street, London WC1E 6BT, UK\\
$^{2}$Dipartimento di Fisica ``G. Galilei'', Universit\`{a} di Padova, via Marzolo 8, I-35131 Padova, Italy\\
$^{3}$Astronomisches Recheninstitut, Zentrum f\"{u}r Astronomie der Universit\"{a}t Heidelberg, Philosophenweg 12, 69120 Heidelberg, Germany\\
$^{4}$Dark Cosmology Centre, Niels Bohr Institute, University of Copenhagen, Juliane Maries Vej 30, DK-2100 Copenhagen, Denmark\\
$^{5}$Aix Marseille Universit\'e, CNRS, LAM (Laboratoire d'Astrophysique de Marseille) UMR 7326, 13388, Marseille, France\\
$^{6}$Exzellenzcluster Universe, Boltzmannstr. 2, 85748 Garching, Germany\\
$^{7}$Ludwig-Maximilians-Universit{\"a}t, Universit{\"a}ts-Sternwarte, Scheinerstr. 1, 81679 M{\"u}nchen, Germany
}
\begin{document}
\label{firstpage}
\pagerange{\pageref{firstpage}--\pageref{lastpage}}
\maketitle

% Abstract of the paper
\begin{abstract}
Tidal gravitational forces can modify the shape of galaxies and clusters of galaxies, thus correlating their orientation with the surrounding matter density field. We study the dependence of this phenomenon, known as intrinsic alignment (IA), on the mass of the dark matter haloes that host these bright structures, analysing the Millennium and Millennium-XXL $N$-body simulations. We closely follow the observational approach, measuring the halo position-halo shape alignment and subsequently dividing out the dependence on halo bias.
We derive a theoretical scaling of the IA amplitude with mass in a dark matter universe, and predict a power-law with slope $\beta_{\mathrm{M}}$ in the range $1/3$ to $1/2$, depending on mass scale.
We find that the simulation data agree with each other and with the theoretical prediction remarkably well over three orders of magnitude in mass, with the joint analysis yielding an estimate of $\beta_{\mathrm{M}} = 0.36^{+0.01}_{-0.01}$. This result does not depend on redshift or on the details of the halo shape measurement. The analysis is repeated on observational data, obtaining a significantly higher value, $\beta_{\mathrm{M}} = 0.56^{+0.05}_{-0.05}$. There are also small but significant deviations from our simple model in the simulation signals at both the high- and low-mass end. We discuss possible reasons for these discrepancies, and argue that they can be attributed to physical processes not captured in the model or in the dark matter-only simulations.
\end{abstract}

% Select between one and six entries from the list of approved keywords.
% Don't make up new ones.

%%%%%%%%%%%%%%%%%%%%%%%%%%%%%%%%%%%%%%%%%%%%%%%%%%

%%%%%%%%%%%%%%%%% BODY OF PAPER %%%%%%%%%%%%%%%%%%

\section{Introduction}
\label{sec:intro}
%In a statistically isotropic universe, one would expect galaxy images not to have preferred orientations; however, we observe that galaxy shapes are locally correlated with the surrounding large-scale structure, and therefore with each other. 
In a statistically isotropic universe, galaxy images have no globally preferred orientation; this, however, does not preclude local correlations of galaxy shapes, with each other and with the large-scale structure.
Several mechanisms have been proposed to explain this, such as the accretion of new material along favoured directions, and the effect of gravitational tidal fields from the surrounding dark matter distribution. In this latter picture, in particular, the shape of luminous structures is affected by tidal interactions by the surrounding dark matter halo, and, consequently, galaxies and clusters also align \citep{Kiesslingetal2015}.

This phenomenon is known as intrinsic alignment (henceforth IA; see \citealt{TroxelIshak2015, Joachimietal2015, Kiesslingetal2015, Kirketal2015} for recent reviews, \citealt{Schmidtetal2015} for its impact in the early-Universe cosmology, and \citealt{L'Huillieretal2017} for its role in the study of different modified gravity and dark energy models): besides carrying information about galaxy formation, intrinsic alignment has an impact on the measurements of the cosmological weak lensing effect, the induced correlations of galaxy image shapes due to gravitational lensing by the mass distribution along the line of sight. IA must therefore be taken into consideration when analysing lensing surveys, such as LSST \citep{LSST2009} and Euclid \citep{Euclid2011}. \citet{Heavensetal2000} and \citet{CroftMetzler2000} first predicted the non-negligible contamination of the weak lensing signal due to correlations in the intrinsic shapes of galaxies, and a number of works afterwards (e.g.\ \citealt{Heymansetal2006, Sembolonietal2008, Kirketal2012}) confirmed that this effect must be accounted for in order not to bias the cosmological inference.
%the results of the observations.

The intrinsic alignment signal has been measured in $N$-body simulations \citep{Heymansetal2006, Kuhlenetal2007, Leeetal2008, Schneideretal2012, Joachimietal2013a, Joachimietal2013b}, in hydrodynamical simulations \citep{Codisetal2015, Velliscigetal2015b, Chisarietal2015, Tennetietal2016, Hilbertetal2017}, and in imaging data \citep{Joachimietal2011, Haoetal2011, Lietal2013, Singhetal2015, vanUitertJoachimi2017}. Results from hydrodynamical simulations and observations, in particular, claimed that massive red galaxies point towards matter overdensities, while blue galaxies do not show any clear sign of alignment \citep{Hirataetal2007, Mandelbaumetal2011}. 

The amplitude of the IA signal has been found to increase with mass in observations. \citet{Haoetal2011}, for example, studied a large sample of galaxy clusters from the SDSS DR7 finding a dependence of the alignment with the mass of the brightest cluster galaxy, and an increasing IA amplitude with luminosity, or the corresponding mass, has been identified for luminous red galaxies (LRGs, \citealt{Joachimietal2011, Singhetal2015}) and for galaxy clusters \citep{vanUitertJoachimi2017}. 

Simulations agree with this picture: using a $512^3$-particle $N$-body simulation, \citet{Jing2002}, for example, found that the alignment increases with mass over three orders of magnitude, up to about $10^{13} M_{\sun}$. Moreover, \citet{Leeetal2008} used data from the Millennium simulation, claiming stronger correlations with higher mass over two mass bins around $10^{12} M_{\sun}$, and \citet{Joachimietal2013b}, using the same set of data, found an increasing trend over the same mass range only for early-type galaxies (whose shapes were assumed to follow those of the underlying haloes), while no dependence on luminosity or particular trend with mass or luminosity for late-type galaxies (whose orientation was determined by the halo spin) was identified. 

These results motivate us to search for a universal relation between the alignment strength and the mass of dark matter haloes. This could lend support to an IA mechanism that successfully explains the trends; moreover, we need to reduce the degrees of freedom in modelling the IA signal to obtain tighter cosmology constraints from lensing observations. To achieve this, we delve into the dependence of the amplitude of the intrinsic alignment signal on the mass of the halo, which is postulated to be the main driver of the shape and orientation of the hosted galaxies (e.g.\ \citealt{Joachimietal2015}). We first derive the expected scaling from the theory, and then test our predictions using data from two $N$-body simulations, mimicking the observational approach in order to be able to straightforwardly compare our results with real data.

We give details about our theoretical model and derive the expected scaling of the IA with halo mass in the tidal alignment paradigm in Sect.~\ref{sec:theory}. We then present the simulations that we use (Sect.~\ref{subsec:sim}), and how we define the shapes of the dark matter haloes they contain (Sect.~\ref{subsec:haloshapes}). In Sect.~\ref{subsec:measurements} we explain how we measure the intrinsic alignment signal, and then (Sect.~\ref{subsec:modelling}) we describe our mass-dependent IA model. We finally show our results and compare them with our theoretical predictions (Sect.~\ref{subsec:simdatard}) and real data (Sect.~\ref{subsec:obsdatard}).

\section{Theoretical background}
\label{sec:theory}
The physical picture of tidal interaction of a self-gravitating system that is in virial equilibrium, with a velocity dispersion $\sigma^2$, such as an elliptical galaxy or a relaxed galaxy cluster, is a distortion of the system's gravitational potential through tidal gravitational forces. The particles of the system remain in virial equilibrium and fill up the distorted potential along an isocontour of the gravitational potential, which results in a change in the shape of the system. This shape modification reflects the magnitude and the orientation of the tidal gravitational fields, and the magnitude of the change in shape depends on how tightly the system is bound.

In isolated virialised systems the Jeans equation applies \citep[see e.g.][]{BinneyTremaine2008}:
\begin{equation}
\frac{1}{\rho}\frac{\partial}{\partial r}(\rho\sigma^2) + \frac{2}{r}\beta_\mathrm{aniso}\sigma^2 = 
-\frac{\partial\Phi}{\partial r} \ ,
\end{equation}
with $\rho$ the particle density, $r$ the distance, $\Phi$ the gravitational potential and $\beta_\mathrm{aniso}$ the anisotropy parameter, which we set to $\beta_\mathrm{aniso}=0$ because we aim to derive only the scaling behaviour of the alignment amplitude. In the case of vanishing anisotropy and a constant velocity dispersion, the Jeans equation can be solved to yield an exponential dependence $\rho\propto\exp(-\Phi/\sigma^2)$ between the density and the gravitational potential.

Gravitational tidal fields generated by the ambient large-scale structure distort the gravitational potential $\Phi$, and we will work in the limit that the distortion is well described by a second-order Taylor-expansion of the potential relative to the centre of the galaxy at $\bmath{r}_0 = 0$. Therefore, the potential relevant for the motion of test particles is given by
\begin{equation}
\Phi(\bmath{r}) \rightarrow
\Phi(\bmath{r}) + \frac{1}{2}\sum_{i,j=1}^{3}\frac{\partial^2\Phi(\bmath{r}_0)}{\partial r_i\partial r_j}r_ir_j \ ,
\end{equation}
with $i,j$ indicating the spatial dimensions. Consequently, the density of particles changes according to
\begin{equation}
\label{eq:density}
\rho = \rho_0
\left(1-\frac{1}{2\sigma^2}\sum_{i,j=1}^{3}\frac{\partial^2\Phi(\bmath{r}_0)}{\partial r_i\partial r_j}r_ir_j\right) \ ,
\end{equation}
with $\rho_0 \propto \exp\left(-\frac{\Phi(\bmath{r})}{\sigma^2}\right)$, under the assumption of a weak distortion such that a Taylor-expansion of the exponential to first order is sufficient. 

The projected ellipticity is calculated through the tensor of second brightness moments \citep{BartelmannSchneider2001}, which we define as:
\begin{equation}
q_{ij} = \int \dif^2x \ \rho_0(\bmath{x}) x_i x_j \ ,
\end{equation}
where the integral is calculated over the plane of the projected sky, with coordinates $x_i,x_j$ and $i,j \in \{1,2\}$. If we consider the distortion of the density described in Eq.~(\ref{eq:density}), we obtain:
\begin{multline}
\label{eq:bridis}
\widetilde{q_{ij}} = \int \dif^2x \ \rho_0(\bmath{x}) \left(1-\sum_{a,b=1}^2\frac{\partial_{a,b}^2\Phi(\bmath{x}_0)}{2\sigma^2} x_a x_b\right) x_i x_j   \\ 
\simeq \int \dif^2 x \ \rho_0(\bmath{x}) x_i x_j - \sum_{a,b=1}^2\frac{\partial_{a,b}^2\Phi}{2\sigma^2} \int \dif^2 x \ \rho_0(\bmath{x}) x_a x_b x_i x_j  \\ \equiv q_{ij} + \psi_{ij} \ ,
\end{multline}
where we switch to a 2D sum since we assume that most relevant are the components of the tidal shear perpendicular to the line of sight, and where the equivalence is not exact due to the approximation that $\partial_{a,b}^2\Phi(\bmath{x}_0) / \sigma^2$ is constant and can be thus taken out of the integral. From this latter equation, we can see that because of the tidal shear fields the second moments of the brightness distribution get a correction $\psi_{ij}$. This term is small in the limit of weak tidal fields, which is characterised by $R^2\partial^2\Phi/\sigma^2\ll 1$ with $R$ being the size of the halo and providing a bound for $r$. 

If we then define the complex ellipticity as in \citet{BartelmannSchneider2001}:
\begin{equation}
\label{eq:ell}
\varepsilon = \frac{q_{11} - q_{22}}{q_{11} + q_{22}+Q} + 2\mathrm{i}\frac{q_{12}}{q_{11} + q_{22}+Q} \ ,
\end{equation} 
with $Q=2\sqrt{q_{11}q_{22}-q_{12}^2}$, and consider the correction obtained in Eq.~(\ref{eq:bridis}), we can write:
\begin{equation}
\label{eq:epsilon}
\widetilde{\varepsilon} = \varepsilon +  \psi_{\epsilon}\ ,
\end{equation} 
with $\varepsilon$ the unperturbed shape as in Eq.~(\ref{eq:ell}), which we assume to be randomly oriented, and
\begin{equation}
\psi_{\epsilon} = \frac{\psi_{11} - \psi_{22}}{q_{11} + q_{22}+Q} + 2i\frac{\psi_{12}}{q_{11} + q_{22}+Q} \ ,
\end{equation}
assuming that $\psi_{ij}$ is a small correction.

%Gravitational tidal fields generated by the ambient large-scale structure distort the gravitational potential $\Phi$, and we work in the limit that the distortion is well-described by a second-order Taylor-expansion relative to the potential minimum at $\bmath{r}_0$,
%\begin{equation}
%\Phi(\bmath{r}) = \Phi(\bmath{r}_0) + \frac{1}{2}\sum_{i,j=1}^{3}\frac{\partial^2\Phi(\bmath{r}_0)}{\partial r_i\partial r_j}(\bmath{r}-\bmath{r}_0)_i(\bmath{r}-\bmath{r}_0)_j \ ,
%\end{equation}
%with $i,j$ indicating the spatial dimensions. Consequently, the density of particles would change according to
%\begin{equation}
%\rho \propto 
%\exp\left(-\frac{\Phi(\bmath{r}_0)}{\sigma^2}\right)
%\times
%\left(1-\frac{1}{2\sigma^2}\sum_{i,j=1}^{3}\frac{\partial^2\Phi(\bmath{r}_0)}{\partial r_i\partial r_j}(\bmath{r}-\bmath{r}_0)_i(\bmath{r}-\bmath{r}_0)_j\right) \ ,
%\end{equation}
%under the assumption of a weak distortion such that a Taylor-expansion of the exponential to first order is sufficient. Physically, this would be the limit of weak tidal fields, which is characterised by $R^2\partial^2\Phi/\sigma^2\ll 1$ with $R$ being the size of the halo, which provides a bound for $\left|\bmath{r}-\bmath{r}_0\right|$.

Therefore, a measurement of the ellipticity of a tidally distorted halo through the second moments of the projected density yields a proportionality to $R^2\partial^2\Phi/\sigma^2$, in accordance with the linear alignment model for elliptical galaxies, where tidal interaction imprints a shape distortion that causes correlations between shapes of neighbouring objects, and the tidally induced ellipticity is proportional to the magnitude of the tidal fields \citep{HirataSeljak2004, Blazeketal2011, Blazeketal2015}. It is worth pointing out that the tidal effect on galaxies is determined by the velocity scale $\sigma^2$, whereas in the case of gravitational lensing the tidal shear is measured in units of the velocity scale $c^2$.

In the case of virialised systems it is possible to relate the velocity dispersion $\sigma^2$ with the size of the object $R$: the virial relationship $\sigma^2 = GM/R$ assumes a proportionality between the specific kinetic energy, $\sigma^2/2$, and the magnitude of the specific potential energy, $GM/R$. With the scaling $M\propto R^3$ one expects $\sigma^2\propto M^{2/3}$, such that $R^2/\sigma^2$ is constant. Therefore, any scaling of the ellipticity with mass is entirely due to the dependence of tidal gravitational fields on the mass scale, and more massive systems are subjected to stronger tidal interactions because of the stronger fluctuations in the tidal fields that they are experiencing.

The variance of tidal shear fields can be inferred from the variance of the matter density by the Poisson equation, $\Delta\Phi = 3\Omega_\mathrm{m}/(2\chi_H^2)\delta$, with $\Omega_\mathrm{m}$ the matter density, $\chi_H=c/H_0$ the Hubble-distance, and $\delta$ the density field. Computing the tidal shear fields $\partial^2\Phi$ shows that they must have the same fluctuation statistics as $\delta$: in Fourier-space, the solution to the Poisson equation is $\Phi \propto \delta/k^2$ with the wave vector $k$, and the tidal shear fields become $k_ik_j\Phi \propto k_ik_j/k^2\:\delta$. Therefore, the power spectrum $P_{\partial^2\Phi}(k)$ of the tidal shear fields is proportional to the power spectrum $P_\delta(k)$ of the density fluctuations.

Consequently, one can derive the variance of the tidal shear fields from the variance of the density fluctuations, i.e.\ from the standard cold dark matter (CDM) power spectrum $P_\delta(k)$. 
Doing so, one can relate a mass scale $M$ to the wave vector $k$ by requiring that the mass $M$ should be contained in a sphere of radius $R$, 
%$M=4\pi\Omega_{\mathrm{m}}\rho_{\mathrm{crit}}R^3/3$,
\begin{equation}
\label{eq:mass}
M=4\pi \Delta \rho_{\mathrm{crit}}R^3/3 = 32\pi^4 \Delta \rho_{\mathrm{crit}}/3k^3
\end{equation}
with $k=2\pi/R$, $\rho_{\mathrm{crit}} = 3H_0^2/8\pi G$ the critical density and $\Delta=200$ an overdensity factor\footnote{Here, the overdensity is relative to the critical density, but note we also consider overdensities relative to the underlying mean matter density. In those cases, we just rescale the mass by a factor of $\Omega_{\mathrm{m}}$, which is the ratio between the mean matter density and the critical density.}.
This defines a scale $k$ in the power spectrum which is proportional to $M^{-1/3}$. This implies that on galaxy and cluster scales, where the CDM-spectrum scales $\propto k^{\gamma}$, with $ \gamma \in [ -3,-2 ]$, one obtains for the standard deviation of the tidal shear field a behaviour $\propto M^{\beta_{\mathrm{M}}}$, with $ \beta_{\mathrm{M}} \in [ 1/3, 1/2]$; in particular, for cluster-size objects ($R=0.5$ Mpc--$1.5$ Mpc), where the non-linear matter power spectrum is proportional to $k^{-2.2}$ \citep{Blasetal2011}, our prediction is $\beta_{\mathrm{M}} \simeq 0.36$. 

We can take this approach a step further: given the premises above, for the amplitude of the intrinsic alignment $A_{\mathrm{IA}}$ we can write
\begin{equation}
\label{eq:secapproach}
A_{\mathrm{IA}} \propto \partial^2\Phi \propto \sqrt{P_{\partial^2\Phi}(k)} \propto \sqrt{P_\delta(k)} \propto \sqrt{P_\delta(M)} \ ,
\end{equation}
where in the last step we assume that Eq.~(\ref{eq:mass}) holds and provides the link between the wave vector $k$ and the mass $M$. This approach overcomes the previous approximation that the power spectrum follows a power-law, and provides a single-parameter model with only a free amplitude; we describe how we test this model in Sect.~\ref{subsec:modelling}.

We emphasise that we are primarily interested in the scaling between ellipticity and tidal shear, which is entirely due to the scale- (or mass-) dependence of the tidal shear field and not due to the internal dynamics of the halo. Predicting the dimensionless constant of proportionality would require many more assumptions in relation to the internal structure of the halo and the size of the system for restricting an otherwise diverging integration.
 
Comparing our work with \citet{Catelanetal2001}, we would like to point out that our virial argument provides indications that the ellipticity of an aligned galaxy is proportional to the tidal shear field and that this proportionality does not depend on mass or redshift. Any scale-dependence of alignments is due to the scale- (or mass-) dependence of the tidal shear fields themselves.

We also emphasise that our model assumes a spherically symmetric, self-gravitating halo in virial equilibrium, with an isotropic velocity distribution with constant dispersion. A consistent solution for such a system would be the singular isothermal sphere with a density profile $\rho\propto 1/r^2$, for which we need to assume a cut-off radius in order to obtain finite values for the quadrupole moments, similarly to \citet{CamelioLombardi2015}. Our virial argument, on the other hand, does not differentiate between the dark matter and stellar components and, in particular, does not assume a potential of the unperturbed galaxy. Gravitational tidal fields would, due to equivalence, act on both components, which should be in virial equilibrium to each other, in the same way: it would not be possible to keep the potential generated by the dark matter component fixed and expose the stellar component to this potential. Furthermore, the potential of the system needs to be consistent with the particle distribution in configuration- and velocity-space, and cannot be chosen independently. These are two points in which we hope to improve the investigation by \citet{CamelioLombardi2015}.

%We compare our very general theoretical predictions with our results in Sect.~\ref{sec:resanddiscuss}.

\section{Data}
\label{sec:data}
\subsection{Simulations}
\label{subsec:sim}
\begin{figure}
	\centerline{
	\includegraphics[scale = 0.305, keepaspectratio]{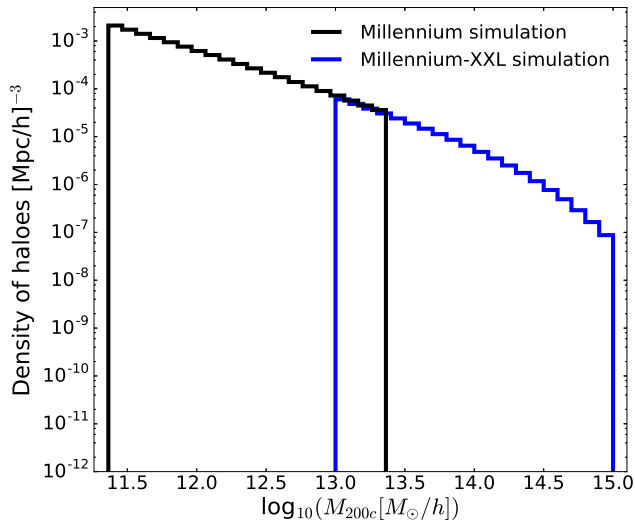}}
	\caption{Histogram showing the number density distribution of the mass of the haloes in the two simulations. Each logarithmic mass bin has a size of $0.1$ dex.
	% the Millennium-XXL catalogue (black line) contains about 6 times more objects than the Millennium catalogue (blue line). 
	In this case, the mass of the halo is defined as $M_{200\mathrm{c}}$, the mass within a region where the density exceeds 200 times the critical density; more details about the chosen mass ranges are presented in Sect.~\ref{subsec:modelling}. In the range where the selected bins overlap, the trend of the density agrees for the two simulations.}
	% Note that in our analysis the Millennium-XXL catalogue is complete only above $10^{13} M_{\sun} h^{-1}$: below this threshold randomly selected haloes were used.}
	\label{fig:histo}
\end{figure}

In this work we consider haloes from two different simulations: 
\begin{enumerate}
\item the \textbf{Millennium simulation (MS)}, first presented in \citet{Springeletal2005}, which uses $2160^3$ dark matter particles of mass $m_{\mathrm{P}}^{\mathrm{MS}} =8.6 \times 10^8 M_{\sun}/h$ enclosed in a 500 Mpc$/h$-side box.
%to sample dark matter haloes and study their growth. 
In particular, we consider 2 of its 64 snapshots, i.e.\ the one at $z = 0$ (snapshot $63$), for our baseline, and the one at $z \simeq 0.46$ (snapshot $49$), to study a potential redshift dependence of our results. Dark matter haloes are identified as in \citet{Joachimietal2013a} and references therein: a simple ``friends-of-friends'' group-finder (FOF, \citealt{Davisetal1985}) is run first to spot virialised structures, followed by the \texttt{SUBFIND} algorithm \citep{Springeletal2001, Springeletal2005} to identify sub-haloes, some of which are then treated as separate haloes if they are only temporarily close to the halo. We only consider haloes with a minimum number of particles $N_{\mathrm{P}} = 300$ \citep{Bettetal2007}.
\item The \textbf{Millennium-XXL simulation (MXXL)}, which samples $6720^3$ dark matter particles of mass $m_{\mathrm{P}}^{\mathrm{MXXL}} = 6.174 \times 10^9 M_{\sun}/h$ confined in a cubic region of $3000$ Mpc$/h$ on a side \citep{Anguloetal2012}. In this case, we consider only one snapshot, at $z = 0$. Haloes are selected using an ellipsoidal overdensity algorithm, as described in \citet{Despalietal2013} and \citet{Bonamigoetal2015}: a traditional spherical overdensity algorithm \citep{LaceyCole1994} gives an initial estimate of the true shape and orientation of the halo, which is then improved by building up an ellipsoid using the previously selected particles. % assuming a minimum number of particles $N_{\mathrm{P}} = 1000$.
\end{enumerate} 
%the shape of the haloes is then derived as for the MS from the eigenvectors and eigenvalues of the mass tensor 
%\begin{equation}
%    \bmath{M}_{\mu \nu} = m_P \sum_{i=1}^{N_P} \frac{r_{i, \mu} r_{i, \nu}}{M_{TOT}} = \frac{1}{N_P} \sum_{i=1}^{N_P} r_{i, \mu} r_{i, \nu}
%	\label{eq:mt}
%\end{equation}
%with $M_{TOT}$ the total mass of the object. Note that the mass tensor (Eq.~\ref{eq:mt}) and the inertia tensor (Eq.~\ref{eq:sit}) are different quantities \citep{Bettetal2007}, but the analysis on the two simulations yields pretty consistent results, as will be shown; also, for this catalogue no ``reduced'' tensor is taken into consideration.

We define the mass of the objects in the catalogues as the mass within a region where the density is 200 times the critical density at the redshift corresponding to the respective snapshot $(M_{200\mathrm{c}})$. Note that for the MS we first convert the halo mass from $M_{\mathrm{Dhalo}}$, as defined in \citet{Jiangetal2014}, to $M_{200\mathrm{c}}$ using the median line in \citet[figure 2]{Jiangetal2014}; this transformation is necessary in order to have a consistent definition of the mass of the haloes in the simulations, but its impact on our results is negligible. The number density distribution of the haloes used in our analysis is shown in Fig.~\ref{fig:histo}.

In the simulations the same set of cosmological parameters is adopted, namely they both assume a spatially flat $\Lambda$CDM universe with the total matter density $\Omega_{\mathrm{m}} = \Omega_{\mathrm{b}} + \Omega_{\mathrm{dm}} = 0.25$, where $\Omega _{\mathrm{b}} = 0.045$ indicates the baryon density parameter and $\Omega_{\mathrm{dm}} = 0.205$ represents the dark matter density parameter, a cosmological constant $\Omega_{\Lambda} = 1 - \Omega_{\mathrm{m}} = 0.75$, the dimensionless Hubble parameter $h = 0.73$, the scalar spectral index $\mathrm{n}_{\mathrm{s}} = 1$, and the density variance in spheres of radius $8 \ \mathrm{Mpc}/h$, $\sigma_8 = 0.9$. 
%All the density parameters are in units of the critical density.

\subsection{Halo shapes}
\label{subsec:haloshapes}
We define the simple inertia tensor\footnote{MS and MXXL use two different tensor definitions to describe the shape, but they result in the same halo ellipticity (see also \citealt{Bettetal2007} for further details).}, whose eigenvalues and eigenvectors describe the shape of the halo, as:
\begin{equation}
    \bmath{M}_{\mu \nu} \propto \sum_{i=1}^{N_{\mathrm{P}}} r_{i, \mu} r_{i, \nu} \ ,
	\label{eq:sit}
\end{equation}
where $N_{\mathrm{P}}$ is the total number of particles within the halo, $\mu, \nu \in \{1, 2, 3\}$, and $\bmath{r}_{i}$ is the vector that indicates the position of the $i-$th particle with respect to the centre of the halo, i.e.\ the location of the gravitational potential minimum, in the reference frame of the simulation box. For the MS only, we also consider a reduced inertia tensor, which is defined as \citep{Pereiraetal2008}:
\begin{equation}
    \bmath{M}_{\mu \nu}^{\mathrm{red}} \propto \sum _{i=1}^{N_{\mathrm{P}}} \frac{r_{i, \mu} r_{i, \nu}}{r_i^2} \ ,
	\label{eq:rit}
\end{equation}
with $r_i^2$ the square of the three-dimensional distance of the i-th particle from the centre of the halo. The reduced inertia tensor is more weighted towards the centre of the halo, and may yield a more reliable approximation of the shape of the galaxy at its centre \citep{Joachimietal2013b, Chisarietal2015}. 

The eigenvectors and eigenvalues define an ellipsoid, which we project onto one of the faces of the simulation box along the $z$-axis: the resulting ellipse is the projected shape of the halo. We proceed as in \citet{Joachimietal2013a} to define the ellipticity $\epsilon$ of the 
% s_{\mathbin{\|}, \mu}
objects, adopting their procedure for early-type galaxies. We denote the eigenvalues of the inertia tensor as $\omega_{\mu}$, and the three eigenvectors as $\bmath{s}_{\mu} = \big \{ s_{x, \mu}, s_{y, \mu}, s_{z, \mu}\big \}^\intercal$, $\mu \in \{1,2,3\}$, where the coordinates refer to the Cartesian system of the simulation box. The projected ellipse is given by the points $\bmath{x}$ which satisfy $\bmath{x}^{\intercal} \bmath{W}^{-1} \bmath{x} =1$, defining a symmetric tensor
\begin{equation}
    \bmath{W}^{-1} = \sum _{\mu=1}^{3} \frac{\bmath{s}_{\perp, \mu} \bmath{s}^\intercal_{\perp, \mu}}{\omega_{\mu}^2} - \frac{\bmath{\kappa} \bmath{\kappa}^{\intercal}}{\alpha^2} \ ,
	\label{eq:symtensor}
\end{equation}
with $\bmath{s}_{\perp, \mu} = \big \{ s_{x, \mu}, s_{y, \mu}\big \}^\intercal$, 
\begin{equation}
   \bmath{\kappa} = \sum_{\mu = 1}^{3} \frac{s_{z, \mu} \bmath{s}_{\perp, \mu}}{\omega_{\mu}^2} \ ,
	\label{eq:kappa}
\end{equation}
and 
\begin{equation}
    \alpha^2=\sum_{\mu = 1}^{3} \left( \frac{s_{z, \mu}}{\omega_{\mu}} \right)^2 \ .
	\label{eq:alpha}
\end{equation}

We compute the two Cartesian components of the ellipticity \citep{BartelmannSchneider2001}
\begin{align}
    	\epsilon_{1} = \frac{W_{11} - W_{22}}{W_{11} + W_{22}+2\sqrt{\det \bmath{W}}}\ , \\ 
           \epsilon_{2} = \frac{2 W_{12}}{W_{11} + W_{22}+2\sqrt{\det \bmath{W}}} \ ,
\end{align}
which we then translate in the radial $(+)$ 
%and cross $(\times)$ 
component, following the IA sign convention\footnote{Our definition of $\epsilon_{+}$, as commonly done in IA works, has an opposite sign with respect to, for example, \citet{BartelmannSchneider2001}.}:
\begin{align}
	\epsilon_{+} = \epsilon_{1} \cos(2\varphi) + \epsilon_{2} \sin(2\varphi)\ ,% \\ 
           %\epsilon_{\times} =  \epsilon_{2} \cos(2\varphi)-\epsilon_{1} \sin(2\varphi) \ , 
\end{align}
where $\varphi$ is the polar angle of the line that connects a halo pair with respect to the $x$-axis, in the reference frame of the simulation box. We show how we use $\epsilon_{+}$ to measure the correlation between halo shapes in the next section.

As a sanity check, we calculate the intrinsic ellipticity dispersion, $\sigma_{\epsilon} = \sqrt{\sum_{i=1}^{N_\mathrm{h}} (\epsilon_1^2+\epsilon_2^2)/N_\mathrm{h}}$ for several mass bins, with $N_\mathrm{h}$ the number of haloes (details about the bins are reported in Sect.~\ref{subsec:modelling} and in Table~\ref{tab:bins}). The magnitude of the dispersion is in good agreement with that of early-type galaxies \citep{Joachimietal2013a} and clusters (0.13 to 0.19 for the cluster samples used in \citealt{vanUitertJoachimi2017}). The dispersion increases with mass, as found in other simulations (e.g.\ \citealt{Despalietal2014, Schrabbacketal2015}), with a small excess for the MXXL samples due to the lower sampling of the halo shapes.

\section{Methodology}
\label{sec:method}
\subsection{Measurements}
\label{subsec:measurements}
To measure the alignment of every dark matter halo with other haloes, we mimic the standard analysis that is usually performed with observations (e.g.\ \citealt{vanUitertJoachimi2017}), i.e.\ we use the halo catalogue as the tracer of the underlying density field. Instead of fitting physical models to the alignment signal and dividing out the galaxy/cluster bias dependence afterwards, though, we take a shortcut: we choose one large bin in the comoving transverse separation $R_{\mathrm{p}}$, and proceed as follows to remove the bias factor dependence.

We first define an estimator as a function of $R_{\mathrm{p}}$ and the line-of-sight distance $\Pi$:
\begin{equation}
    \hat{\xi}_{\mathrm{g+}}(R_{\mathrm{p}}, \Pi) = \frac{S_+ D}{D D} \ ,
	\label{eq:xigphat}
\end{equation}
where $S_+ D$ represents the raw correlation between halo shapes ($\epsilon_{+}$) and the density sample, and $D D$ the number of halo shape-density pairs. Per default, the halo shape sample is also used as the density sample. We then integrate along the line of sight to obtain the total projected intrinsic alignment signal:
\begin{equation}
   \hat{w}_{\mathrm{g+}} (R_{\mathrm{p}}) = \int _{-\Pi_{\mathrm{max}}}^{\Pi_{\mathrm{max}}} \dif \Pi \ \hat{\xi}_{\mathrm{g+}}(R_{\mathrm{p}}, \Pi) \ .
	\label{eq:wgphat}
\end{equation}
Throughout this work, we adopt $\Pi_{\mathrm{max}} = 60 $ Mpc$/h$, a value large enough not to miss part of the signal, but small enough not to pick up too much noise. 
We describe the intrinsic alignment signal by simplifying the model in Eq.~5 of \citet{vanUitertJoachimi2017}, namely we assume:
\begin{equation}
    w_{\mathrm{g+}} (R_{\mathrm{p}}, M)=A_{\mathrm{IA}} (M) \ b_{\mathrm{h}} (M)\ w_{\mathrm{\delta +}}^{\mathrm{model}} (R_{\mathrm{p}}) \ , 
	\label{eq:wgp}
\end{equation}
with $A_{\mathrm{IA}} (M)$ the amplitude of the intrinsic alignment signal, $b_\mathrm{h} (M)$ the halo bias, and $w_{\mathrm{\delta} \mathrm{+}}^{\mathrm{model}} (R_{\mathrm{p}})$ a function in which we include the dependence on $R_{\mathrm{p}}$. In the tidal alignment paradigm, $w_{\mathrm{\delta} \mathrm{+}}^{\mathrm{model}}$ is independent of halo mass, since it is fully determined by the properties of the dark matter distribution, assuming that any mass dependence of the response of a halo shape to the tidal gravitational field is captured by $A_{\mathrm{IA}}(M)$. 

We evaluate the expression in Eq.~(\ref{eq:wgp}) in the interval which covers $10\ \mathrm{Mpc}/h < R_{\mathrm{p}} < 20 \ \mathrm{Mpc}/h$, denoted by $R_{\mathrm{p}}^*$, to remove the dependence on $R_{\mathrm{p}}$; in other words, we define the halo-pair weighted average in $R_{\mathrm{p}}^*$ as:
\begin{equation}
    w_{\mathrm{g+}}^* (M) \equiv \langle w_{\mathrm{g+}} (R_{\mathrm{p}}, M) \rangle _{R_{\mathrm{p}}^*} \ .
	\label{eq:wgpmass}
\end{equation}
We choose this particular interval for two reasons. First, $10$ Mpc$/h$ is well above the minimum scale usually adopted in observational papers, below which the bias becomes non-linear \citep{Tasitsiomietal2004, Mandelbaumetal2006}. Second, we gain little from extending the range beyond 20 Mpc/h, as the signal-to-noise ratio of the IA signal is small at larger scales. As a sanity check, we repeat our analysis in an extended interval, which covers $6\ \mathrm{Mpc}/h < R_{\mathrm{p}} < 30 \ \mathrm{Mpc}/h$, without significantly changing our results. In a very broad bin the effective radial weighting of halo pairs could vary significantly across the mass range considered; hence we prefer the narrower bin as our default.

In order to constrain $b_{\mathrm{h}} (M)$ in Eq.~(\ref{eq:wgp}), we use the LS \citep{LandySzalay1993} estimator to calculate the clustering signal:
\begin{equation}
    \hat{\xi}_{\mathrm{gg}}(R_{\mathrm{p}}, \Pi) = \frac{DD -2DR + RR}{RR} \ ,
	\label{eq:xigghat}
\end{equation}
where $DD$ represents the number of halo pairs, $DR$ the number of halo-random point pairs, and $RR$ the number of random point pairs. To measure $DR$ and $RR$, we generate random catalogues that contain objects uniformly distributed between the minimum and maximum value of the $x$, $y$ and $z$ coordinates of each simulation sub-box\footnote{For further information about the sub-boxes, see Sect.~\ref{subsec:modelling}.}. These catalogues normally are three times denser, but in some cases, when a sub-box encloses very few objects, we switch to random catalogues which are ten times denser. We always use Eq.~(\ref{eq:xigghat}) since we re-normalise the estimators according to the sample size. We then integrate along the line of sight to obtain the total projected clustering signal:
\begin{equation}
   \hat{w}_{\mathrm{gg}} (R_{\mathrm{p}}) = \int _{-\Pi_{\mathrm{max}}}^{\Pi_{\mathrm{max}}} \dif \Pi \ \hat{\xi}_{\mathrm{gg}}(R_{\mathrm{p}}, \Pi) \ .
	\label{eq:wgghat}
\end{equation}

We describe the clustering signal with a simple model:
\begin{equation}
    w_{\mathrm{gg}} (R_{\mathrm{p}}, M)=b_\mathrm{h}^2 (M)\ w_{\mathrm{\delta} \mathrm{\delta}}^{\mathrm{model}} (R_{\mathrm{p}}) - C_{\mathrm{IC}} \ , 
	\label{eq:wgg}
\end{equation}
with $w_{\mathrm{\delta} \mathrm{\delta}}^{\mathrm{model}} (R_{\mathrm{p}})$ a function in which we include the dependence on $R_{\mathrm{p}}$ \citep[equation 9]{vanUitertJoachimi2017}, and $C_{\mathrm{IC}}$ the integral constraint, which accounts for the offset due to the restricted area of the simulation box. We assess that this correction is negligible for the MXXL by measuring that the clustering signals are unaltered if we normalise the random catalogues with respect to the whole simulation box rather than the sub-boxes. We estimate the integral constraint for the MS by selecting an identical mass bin in both simulations $(10^{13} M_{\odot}/h - 10^{13.5} M_{\odot}/h)$, and calculating $w_{\rm{gg}}$ as a function of $R_{\rm{p}}$. Assuming that the MXXL signal has a negligible $C_{\mathrm{IC}}$, we find the optimal $C_{\mathrm{IC}}$ for the MS signal via weighted least squares fitting of the MS $w_\mathrm{gg}$ to the MXXL $w_\mathrm{gg}$. We then deduce the individual integral constraints for each MS mass bin by assuming that they scale with $b_\mathrm{h}^2(M)$. Note that, since we are dealing with dark matter clustering, $w_{\mathrm{\delta} \mathrm{\delta}}^{\mathrm{model}} (R_{\mathrm{p}})$ does not depend on the mass of the halo. Again, we average the previous expression in $R_{\mathrm{p}}^*$, obtaining:
\begin{equation}
    w_{\mathrm{gg}}^* (M) \equiv \langle w_{\mathrm{gg}} (R_{\mathrm{p}}, M) \rangle _{R_{\mathrm{p}}^*} \ .
	\label{eq:wgpmass}
\end{equation}

In observational analyses, the halo bias factor is usually measured from a fit to $w_{\mathrm{gg}}$ and then included in the modelling of $w_{\mathrm{g+}}$; equivalently, to remove the mass dependence of the halo bias $b_{\mathrm{h}} (M)$, we define:
\begin{equation}
    r_{\mathrm{g+}} (M)=\frac{w_{\mathrm{g+}}^* (M)}{\sqrt{w_{\mathrm{gg}}^*(M)+C_{\mathrm{IC}}}} = \frac{A_{\mathrm{IA}} (M) \langle w_{\mathrm{\delta} \mathrm{+}}^{\mathrm{model}} (R_{\mathrm{p}}) \rangle _{ R_{\mathrm{p}}^*} }{\sqrt{\langle w_{\mathrm{\delta} \mathrm{\delta}}^{\mathrm{model}} (R_{\mathrm{p}})\rangle _{R_{\mathrm{p}}^*}}} \propto A_{\mathrm{IA}} (M) \ ,
	\label{eq:rg+}
\end{equation}
where we assume that the clustering signal $w_{\mathrm{gg}}^*(M)$ is positive (see Sect.~\ref{subsec:modelling} for further discussion). 
%The equivalence is not exact due to the small correction of the integral constraint $C_{\mathrm{IC}}$ in the MS clustering signal. 
We stress that, under our assumptions, $r_{\mathrm{g+}}$ depends only on the mass of the halo M.

\subsection{Modelling}
\label{subsec:modelling}
%The goal of this paper is to study the dependence on the mass of the amplitude $A_{\mathrm{IA}} (M)$ by studying the quantity $r_{\mathrm{g+}} (M)$.
The goal of this paper is to study the halo mass dependence of the intrinsic alignment amplitude $A_{\mathrm{IA}} (M)$, which we constrain by fitting $r_{\mathrm{g+}} (M)$.
%\begin{equation}
%A_{\mathrm{IA}} (M)\propto M^{\beta_{\mathrm{M}}} \ , 
%	\label{eq:aia}
%\end{equation}

We select the haloes from the catalogues described in Sect.~\ref{subsec:sim} in $n_\mathrm{M} = 4$ logarithmic mass bins, each extending over two orders of magnitude, between $10^{11.36} M_{\sun}/h$ and $10^{13.36} M_{\sun}/h$ for the MS and between $10^{13} M_{\sun}/h$ and $10^{15} M_{\sun}/h$ for the MXXL (see Table~\ref{tab:bins}). We split them in $N_\mathrm{sub} = 3^3 = 27$ sub-boxes based on their positions inside the cube of the respective simulation, and calculate $w_{\mathrm{g+}}^*$ and $w_{\mathrm{gg}}^*$ for each sub-sample by replacing the integrals in Eq.~(\ref{eq:wgphat}) and~(\ref{eq:wgghat}) with a sum over 20 line-of-sight bins, each $2\Pi_{\mathrm{max}}/ 20 \ = 6 \ \mathrm{Mpc}/h$ wide, defining the line of sight along the $z$-axis. 

\begin{table*}
	\centering
	\caption{Mass range, number of haloes $N_\mathrm{h}$, mean ellipticity $\langle \epsilon \rangle$ and ellipticity dispersion $\sigma_{\epsilon}$ for each of the 4 bins of both the Millennium and the Millennium-XXL simulations. The masses are defined as $M_{200\mathrm{c}}$, the mass within a region where the density exceeds 200 times the critical density.}
	\label{tab:bins}
	\tabcolsep=0.25cm
	\begin{tabular}{c||cccc} % four columns, alignment for each
		\hline \hline
		\ & Mass range $[M_{\sun}/h]$ & $N_\mathrm{h}$ & $\langle \epsilon \rangle$ &  $\sigma_{\epsilon}$  \\
		\hline
		\multirow{4}{*}{Millennium simulation} & $10^{11.36}-10^{11.86}$ & $916109$ &$0.13$& $0.15$  \\ \ & $10^{11.86}-10^{12.36}$ & $328364$ &$0.13$& $0.15$  \\ \ & $10^{12.36}-10^{12.86}$ & $113917$ &$0.14$& $0.16$ \\ \ & $10^{12.86}-10^{13.36}$ & $37511$ &$0.16$& $0.18$ \\
		\hline 
		\multirow{4}{*}{Millennium-XXL simulation} & $10^{13}-10^{13.5}$ & $5450501$ &$0.19$& $0.22$  \\ \ & $10^{13.5}-10^{14}$ & $1618707$ &$0.21$& $0.24$  \\ \ & $10^{14}-10^{14.5}$ & $370911$ &$0.23$& $0.26$ \\ \ & $10^{14.5}-10^{15}$ & $48714$ &$0.25$& $0.28$ \\
		\hline \hline 
		\end{tabular}
\end{table*}

We estimate the covariance matrix on the mean of the sub-boxes:
\begin{equation}
     \bmath{C}_{\mu \nu} = \frac{1}{N_\mathrm{sub}(N_\mathrm{sub}-1)} \sum_{j = 1}^{N_\mathrm{sub}} (d_{j, \mu} - \overline{d}_{\mu})(d_{j, \nu} - \overline{d}_{\nu}) \ ,
	\label{eq:covariance}
\end{equation}
with $\mu, \nu \in \{1, \dotso, n_\mathrm{M}\}, \overline{d}_{\mu} = \frac{1}{N_\mathrm{sub}} \sum_{j=1}^{N_\mathrm{sub}} d_{j, \mu},$ and $d_{j, \mu}= r_{\mathrm{g+}}(M)$ for each sub-box and each mass bin, as defined in Eq.~(\ref{eq:rg+}). We then invert the covariance matrix and correct the bias on the inverse to obtain an unbiased estimate of the precision matrix, given by:
\begin{equation}
     \bmath{C}^{-1}_{\mathrm{unbiased}} = \frac{N_\mathrm{sub} - n_\mathrm{M} -2}{N_\mathrm{sub}-1} \  \bmath{C}^{-1} \ ,
	\label{eq:precunbiased}
\end{equation}
where $N_\mathrm{sub} > n_\mathrm{M}+2$ clearly holds \citep{Kaufman1967, Hartlapetal2007, Tayloretal2013}. We assess that this procedure does not introduce any systematic errors in the covariance assigned to our measurements by calculating our signals in $N_\mathrm{sub}$ MS-size boxes selected from the MXXL, and then observing that the results do not significantly differ from the MS signals.

The choice of $n_\mathrm{M}$ and $N_\mathrm{sub}$ is constrained by several factors: first of all, if $N_\mathrm{sub}$ is too large, the single values of $w_{\mathrm{gg}}^*$ (and $w_{\mathrm{g+}}^*$) tend to fluctuate around the mean, thus increasing the error bar and sometimes dropping below 0, which is unacceptable for our choice of $ r_{\mathrm{g+}} (M)$; see Eq.~(\ref{eq:rg+}). Furthermore, we want $n_\mathrm{M}$ to be large enough to be capable of displaying the trend of the signals along the whole mass range chosen. Finally, we need to take $n_\mathrm{M} \ll N_\mathrm{sub}$ to avoid divergences related to the fact that we estimate the covariance from a finite number of samples \citep{Tayloretal2013}.

As a first approach, which we will denote as ``power-law'' approach, we adopt the following model for $r_{\mathrm{g+}} (M)$:
\begin{equation}
    r_{\mathrm{g+}} (M) = A \cdot  \left( \frac{M}{M_{\mathrm{p}}} \right)^{\beta_{\mathrm{M}}} \ ,
	\label{eq:modelrg+}
\end{equation}
with $A$ a generic amplitude which we will treat as a nuisance parameter, $M_{\mathrm{p}} = 10^{13.5} M_{\sun}/h_{70}$ a pivot mass, and $\beta_{\mathrm{M}}$ a free power-law index, which we intend to compare with the value predicted in Sect.~\ref{sec:theory}.
We perform a likelihood analysis over the data to infer the posteriors of $A$ and $\beta_{\mathrm{M}}$. According to Bayes' theorem, if $\bmath{d}$ is the vector of the data and $\bmath{p}$ the vector of the parameters, 
\begin{equation}
    P(\bmath{p} | \bmath{d}) \propto P(\bmath{d} | \bmath{p}) \ P(\bmath{p}) \propto e^{-\frac{1}{2} \chi ^2} P(\bmath{p}) \ ,
	\label{eq:bayes}
\end{equation}
with $P(\bmath{p} | \bmath{d})$ the posterior probability, $P(\bmath{d} | \bmath{p})$ the likelihood function, $P(\bmath{p})$ the prior probability and $\chi ^2 = (\bmath{d} - \bmath{m})^T \bmath{C}^{-1} (\bmath{d} - \bmath{m})$, with $\bmath{m}$ the vector of the model and $\bmath{C}^{-1}$ the precision matrix, the inverse of the covariance matrix $\bmath{C}$. We assume uninformative flat priors in the fit with ranges $\log_{10} \frac{A}{({\mathrm{Mpc}/h})^{1/2}} \in [-1.3;-0.6]$ and $\beta_{\mathrm{M}} \in [0.2;0.7]$.
%which generously include the range around the best-fit values. 

As a second approach, which we will denote as ``power spectrum'' approach, we use the CLASS (Cosmic Linear Anisotropy Solving System, \citealt{Blasetal2011}) algorithm to generate a non-linear matter power spectrum using the simulation cosmology. We then use Eq.~(\ref{eq:mass}) to convert the wave number to the corresponding mass $M$, and calculate the square root of the values of the power spectrum $P_\delta(M)$, as in Eq.~(\ref{eq:secapproach}). We adopt the following model for $r_{\mathrm{g+}} (M)$:
\begin{equation}
 r_{\mathrm{g+}} (M) = A_\mathrm{PS} \cdot  \sqrt{P_\delta(M)} \ ,
 \label{eq:secapproachmodel}
\end{equation}
with $A_\mathrm{PS}$ a generic amplitude which we will treat as a nuisance parameter.

%We generate $N_\mathrm{PS}=1000$ models as in Eq.~\ref{eq:secapproachmodel}, varying the amplitude $\log_{10}{\big(A_\mathrm{PS}/h\mbox{Mpc}^{-1}\big)}$ between $-2.5$ and $-0.5$, and then we calculate for all of them the $\chi^2$:
%\begin{multline}
%\chi^2 (A_\mathrm{PS})=\sum_{i,j=1}^{n_\mathrm{PS}} \left(r_{\mathrm{g+}} (M_i)- A_\mathrm{PS} \sqrt{P_\delta(M_i)}\right) \times \\ \times \bmath{C}^{-1}_{ij} \left(r_{\mathrm{g+}} (M_j)- A_\mathrm{PS} \sqrt{P_\delta(M_j)}\right)
%\end{multline}
%where the inverse of the covariance matrix is corrected as in Eq.~\ref{eq:precunbiased}, and $n_\mathrm{PS}=2 n_\mathrm{M}$ is the total number of bins, to include both simulations in the analysis.

To be able to directly compare our results with those in \citet[figure 7]{vanUitertJoachimi2017}, we further convert the halo mass definition from $M_{\mathrm{200c}}$ to $M_{\mathrm{200m}}$, defined as the mass enclosed in a region inside of which the density is 200 times the mean density at the redshift corresponding to the respective snapshot.

\section{Results and discussion}
\label{sec:resanddiscuss}
\subsection{Simulation data}
\label{subsec:simdatard}
\begin{figure}
	\centerline{
	\subfloat[$w_{\mathrm{g+}}$ and $w_{\mathrm{gg}}$ for the Millennium simulation.]
	{\includegraphics[scale = 0.7, keepaspectratio]{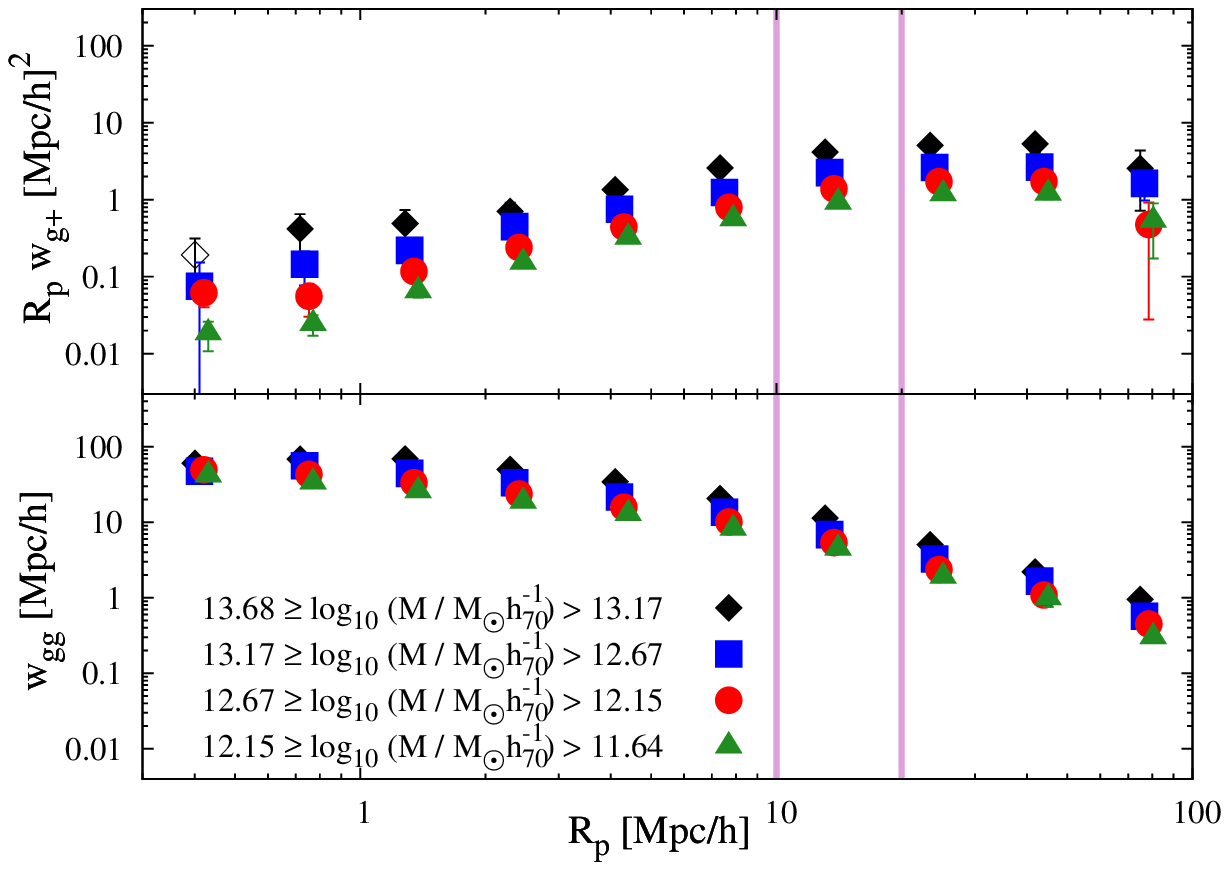}
	\label{wgpwggrp1}}}
	\centerline{
	\subfloat[$w_{\mathrm{g+}}$ and $w_{\mathrm{gg}}$ for the Millennium-XXL simulation.]	
	{\includegraphics[scale = 0.7, keepaspectratio]{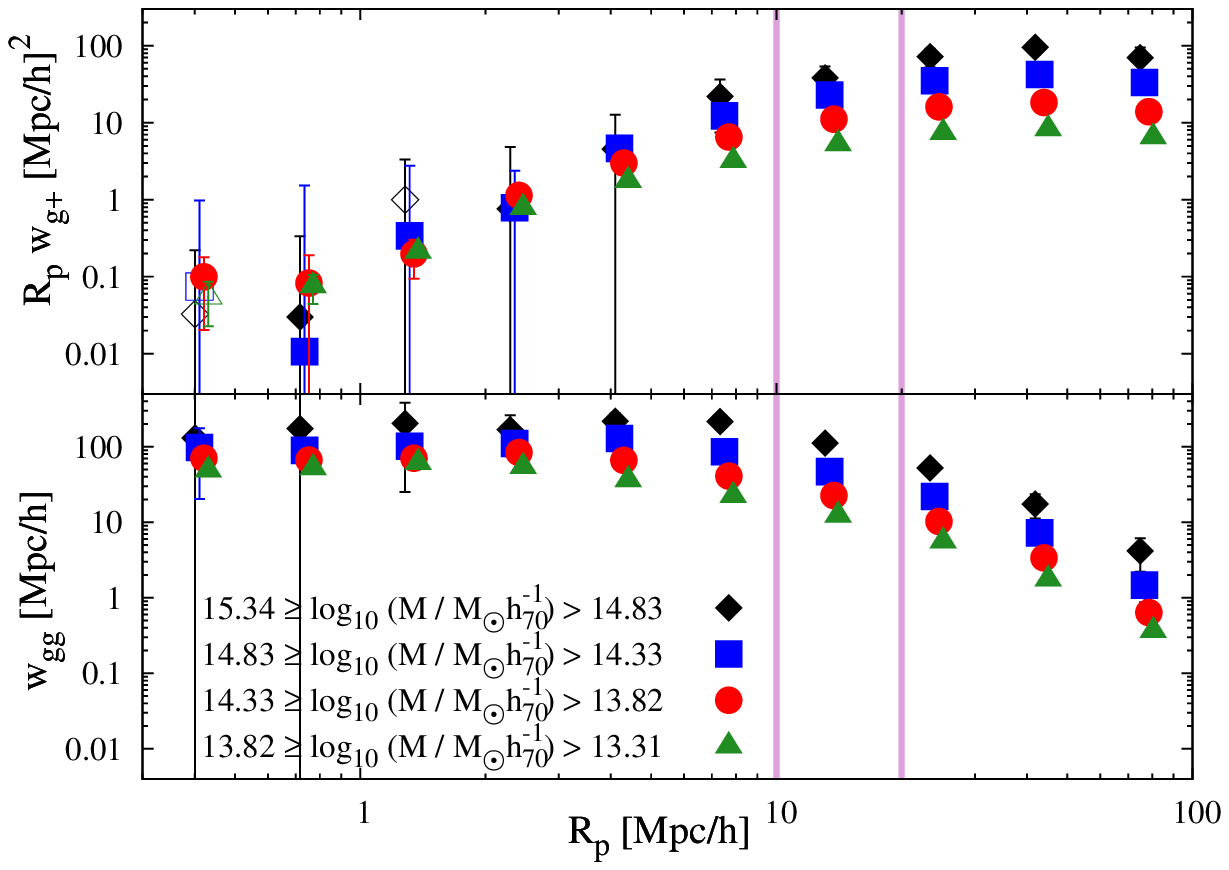}
	\label{wgpwggrp2}}}
	\caption{The intrinsic alignment signal $w_{\mathrm{g+}}$ and the clustering signal $w_{\mathrm{gg}}$ as a function of the comoving transverse separation $R_{\mathrm{p}}$ at $z=0$ for \protect\subref{wgpwggrp1} the Millennium and \protect\subref{wgpwggrp2} the Millennium-XXL simulation. The pink lines indicate the $10 < R_{\mathrm{p}} / \ h^{-1}\mathrm{Mpc} < 20 $ interval, the range used to model the signal. The mass ranges are displayed considering $M_{\mathrm{200m}}$ as the mass of the halo.
	%, and correspond to the second and the fourth bin of our division, explained in detail in Sect.~\ref{subsec:modelling}. The trend for all the mass bins in shown in Appendix~\ref{app:plot}. 
	In the graph, points are slightly horizontally shifted, so that they do not overlap; negative values are displayed in absolute value with open symbols of the same colour. An increasing trend with mass is clear in each panel separately, and comparing the two panels as well.} %We note that $w_{\mathrm{gg}}$ becomes negative at low $R_{\mathrm{p}}$ only for the highest mass bin of the MXXL, where the number density of haloes is lower than in the other bins, as shown in Fig.~\ref{fig:histo}.}
	\label{fig:wgpwggrp}
\end{figure}

The trend of $w_{\mathrm{g+}}$ and $w_{\mathrm{gg}}$ with $R_{\mathrm{p}}$ at $z=0$ for the four mass bins for each simulation is shown in Fig.~\ref{fig:wgpwggrp}.
% Note that even though in the chosen interval $w_{\mathrm{gg}}$ is not always positive within the error bar, our choice of $n_\mathrm{M}$ and $N_\mathrm{sub}$ and our large-bin average ensure that Eq.~\ref{eq:rg+} always returns a real value. 
The points shown in Fig.~\ref{fig:wgpwggrp} are the arithmetic mean of the $N_\mathrm{sub}$ values for each mass bin, while the error bars are the standard deviation of the mean values. The overall behaviour of $w_{\mathrm{g+}}$ and $w_{\mathrm{gg}}$ agrees with previous works \citep{Joachimietal2011, vanUitertJoachimi2017}; in particular, we clearly detect positive intrinsic alignment in all samples, implying that the projected ellipticities of dark matter haloes tend to point towards the position of other haloes. Also, it is worth noting that both signals increase with increasing mass.

\begin{figure}
	\centerline{
	\includegraphics[scale = 0.7, keepaspectratio]{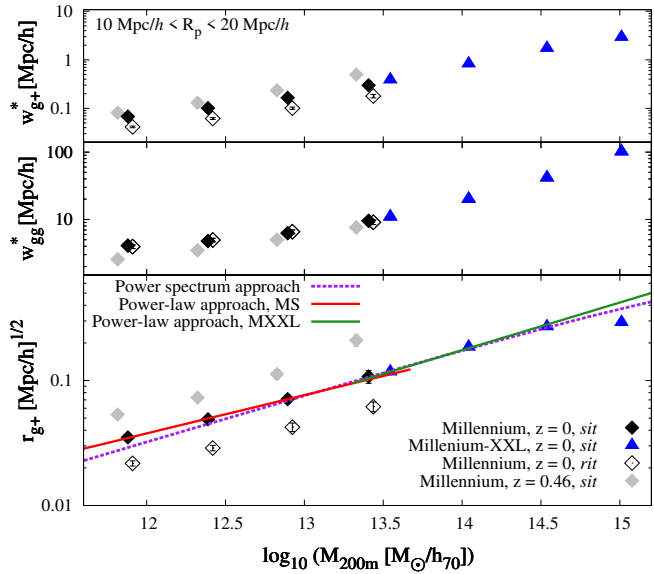}}
	\caption{The intrinsic alignment signal $w_{\mathrm{g+}}^*$, the clustering signal $w_{\mathrm{gg}}^*$ and $r_{\mathrm{g+}}$ as defined in Eq.~(\ref{eq:rg+}) as a function of halo mass $M_{\mathrm{200m}}$ for the Millennium and the Millennium-XXL simulations, calculated as a weighted average in the interval which covers $10\ \mathrm{Mpc}/h < R_{\mathrm{p}} < 20 \ \mathrm{Mpc}/h$. The label \textit{sit} stands for simple inertia tensor, while \textit{rit} means reduced inertia tensor; note that for the MXXL data only the simple inertia tensor is available, as mentioned in Sect.~\ref{subsec:haloshapes}. The points are not placed at the midpoint of the bin, but at the value corresponding to the arithmetic mean of the mass of the objects. The red line and the green line represent the best-fit lines for the MS and MXXL likelihood analyses, respectively, which are drawn using the parameters reported in Table~\ref{tab:param}, while the purple dashed line represents the best-fit line for the ``power spectrum'' model. With both approaches, we exclude the highest-mass bin from the fit. Points showing the results from the \textit{rit} choice are horizontally shifted by a small amount, so that they do not overlap with the corresponding \textit{sit} dots.}
	\label{fig:vsmass}
\end{figure}

We then study the dependence of $w_{\mathrm{g+}}^*, w_{\mathrm{gg}}^*$ and $r_{\mathrm{g+}}$ on the mass of the halo. As one can see from Fig.~\ref{fig:vsmass}, despite the use of two different halo finders, the MS and the MXXL follow the same trend, and yield consistent results in the small mass range where they overlap; furthermore, all three $w_{\mathrm{g+}}^*, w_{\mathrm{gg}}^*$ and $r_{\mathrm{g+}}$ increase with increasing mass. In Fig.~\ref{fig:vsmass} we also include, for the Millennium simulation only, two more results: grey diamonds represent the signal from the objects at redshift $z = 0.46$, while open black diamonds represent the signal from the objects at $z = 0$ obtained using the reduced inertia tensor (\textit{rit}, as in Eq.~\ref{eq:rit}) to measure the shapes of the haloes. We find that the use of the reduced inertia tensor leads to lower alignment signals, and that these signals increase with increasing redshift, as found in \citet{Joachimietal2013b}. Importantly, we do not see any substantial deviations from the general trend with mass for these alternative measurements, suggesting that the details about redshift and halo-shape definition are absorbed in the nuisance parameter A.

\begin{figure*}
	%\centerline{
	%\subfloat[Posterior analysis for the Millennium simulation.]
	%{\includegraphics[scale = 0.7]{inference_millennium.eps}
	%\label{fig:postsim1}}
	%\subfloat[Posterior analysis for the Millennium-XXL simulation.]	
	%{\includegraphics[scale = 0.7]{inference_mxxl.eps}
	%\label{fig:postsim2}} }
	\centerline{	
	\subfloat[Single and joint likelihood analysis for the simulations.]
	{\includegraphics[scale = 0.7, keepaspectratio]{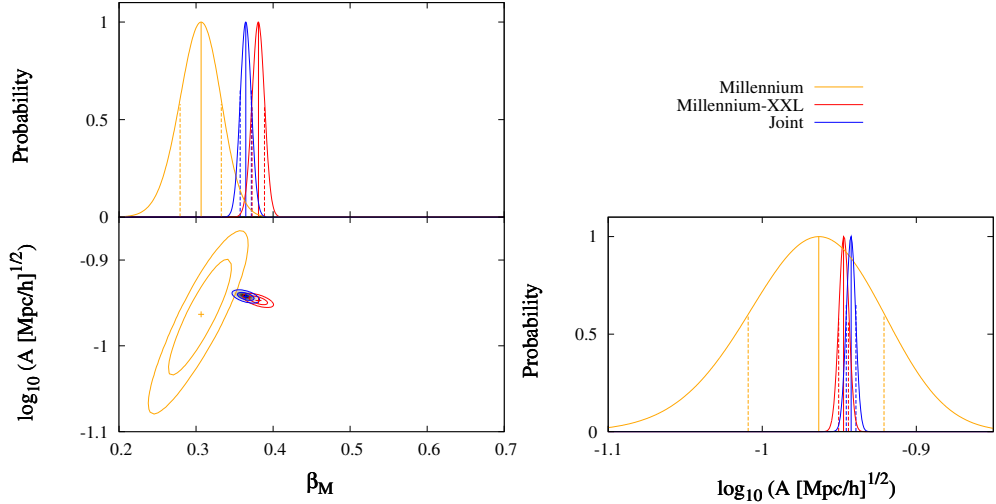}
	\label{fig:postsim3}}}
	\centerline{
	\subfloat[Likelihood analysis for the real data.]
	{\includegraphics[scale = 0.7, keepaspectratio]{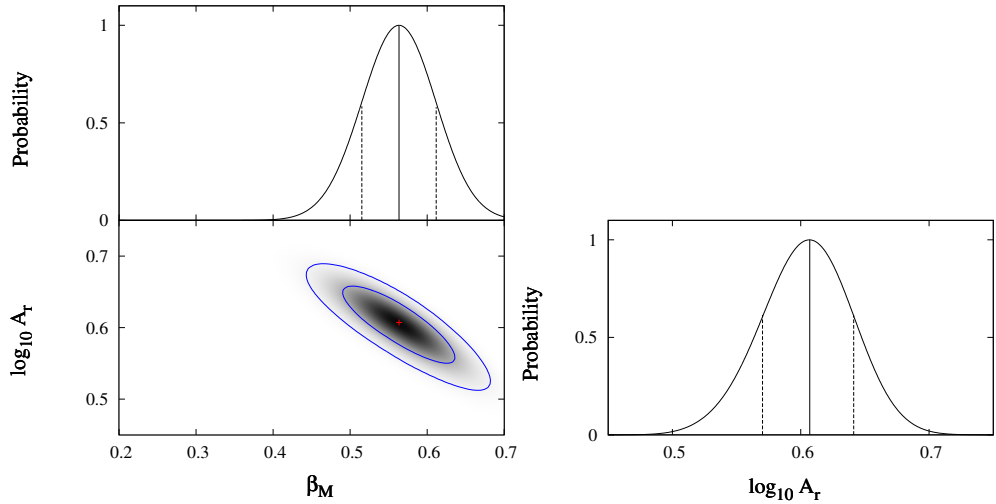}
	\label{fig:postreal}} }
	\caption{Posterior on the fit parameters of the IA model, obtained for \protect\subref{fig:postsim3} the Millennium simulation, the Millennium-XXL simulation, joint MS and MXXL, and \protect\subref{fig:postreal} real data. Note that the ranges of the prefactors are quite different. The bottom-left graph in the top set of panels shows the contour lines and the 2-D greyscale posterior for the real data, while the bottom-left graph in the lower set of panels shows the contour lines of the 2-D posteriors for all the simulations (single and joint), but the 2-D greyscale posterior for the joint analysis only. All other sub-panels show the marginalized 1-D posterior normalized to a peak amplitude of 1. Contour lines enclose the 68\% and 95\% confidence intervals, crosses and vertical solid lines indicate the best-fitting values, while dashed lines represent the $1$-$\sigma$ confidence interval. We note that the MS returns larger error bars, but the results are consistent for the two catalogues, while real data yield a value of the slope which is incompatible with the one from the joint analysis. The different degeneracy for the parameters in the two simulations is due to the choice of the pivot mass $M_{\mathrm{p}} = 10^{13.5} M_{\sun}/h_{70}$: for the MS, it is at the upper end of the mass scale, while for the MXXL it is at the lower end. The mean values and 68\% confidence intervals of $A$, $A_r$ and $\beta_{\mathrm{M}}$ are listed in Table~\ref{tab:param}.}
	\label{fig:post}
\end{figure*}

We proceed by showing the results of the likelihood analysis of the ``power-law'' approach described in Sect.~\ref{subsec:modelling}: 
%Fig.~\ref{fig:post}\protect\subref{fig:postsim1} and Fig.~\ref{fig:post}\protect\subref{fig:postsim2} display the outcomes of the separated analysis on the two catalogues, while 
Fig.~\ref{fig:post}\protect\subref{fig:postsim3} shows the results from the individual datasets and from the joint analysis of the two simulations, obtained by multiplying the likelihood functions and assuming the same flat priors on the parameters. In the MXXL and in the joint fit, we exclude the highest-mass point, since it clearly does not follow our model predictions. We test and discuss possible reasons for this deviation below.

The most stringent bounds come from the MXXL, while the MS yields larger errors on the parameters, albeit consistent with the results of the MXXL within $3\sigma$. The joint analysis returns a value for the slope compatible with $\beta_{\mathrm{M}} = 1/3$, which agrees with the predictions made in Sect.~\ref{sec:theory} for a DM-only universe; in particular, the tightest constraints are obtained with cluster-size objects, for which we predicted $\beta_{\mathrm{M}} \simeq 0.36$, in agreement with our MXXL high-mass results $\big( \beta_{\mathrm{M}} = 0.38^{+0.01}_{-0.01} \big)$. Moreover, we find that neither the inertia tensor definition nor the chosen redshift for our default analysis have significant impact on our conclusions for $\beta_{\mathrm{M}}$. %although we note that the alignment amplitude increases with $z$.
We report all the best-fit values, together with their respective errors and reduced $\chi^2$, in Table~\ref{tab:param}.

In Fig.~\ref{fig:vsmass} we also show the result of our ``power spectrum'' approach: we choose $\Delta = 200\Omega_{\mathrm{m}}$ in Eq.~(\ref{eq:mass}) to be consistent with the definition of the mass, and in the fit we exclude the highest-mass result.
%therefore $n_\mathrm{PS}=2 n_\mathrm{M} - 1 = 7$. 
Our best-fit line corresponds to $\log_{10}{\big(A_\mathrm{PS}/h\mbox{Mpc}^{-1}\big)} =  -1.54$ and $\chi^2 / \mbox{d.o.f.} = 6.16$. In this case, the large value of the reduced $\chi^2$ is largely driven by the two lowest-mass points.

\begin{figure}
	\centerline{
	\includegraphics[scale = 0.7, keepaspectratio]{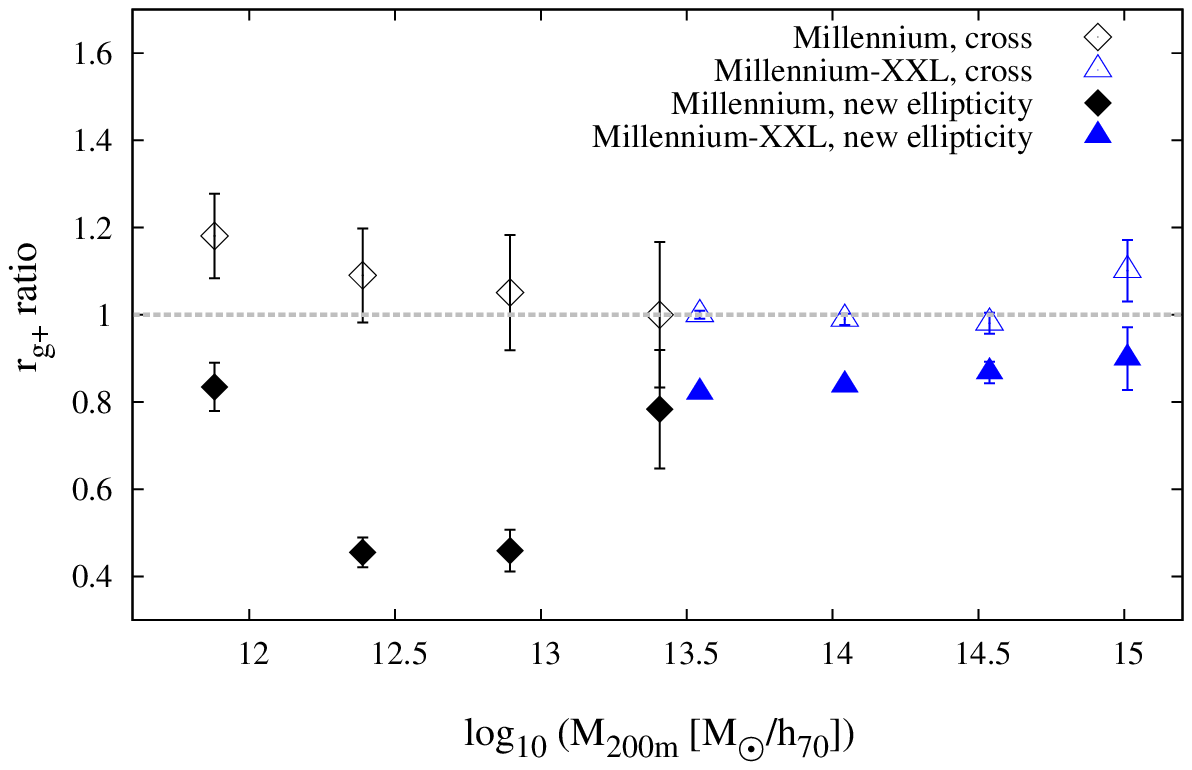}}
	\caption{Ratio of $r_{\mathrm{g+}}$ as a function of halo mass for two sets of signals, with respect to the results of our main analysis shown in Fig.~\ref{fig:vsmass} as black and blue diamonds. Open diamonds are derived from the signals obtained with constant density tracers, which we measure to assess our treatment of the halo bias. Filled diamonds are derived from the signals obtained with all ellipticities rescaled to the mean in each sample (see Eq.~\ref{eq:epsilonprime} and~\ref{eq:epsilonprimesecond}), and demonstrate that the intrinsic halo ellipticity affects the mass scaling of the alignment signal. We neglect any correlations between the $r_\mathrm{g+}$ measurements in estimating the error on the ratio.}
	\label{fig:ratio}
\end{figure}

Although our simple model impressively reproduces the general mass trend, the bad goodness of the fit in most cases indicates that there are deviations which are significant beyond the very small statistical error bars of our simulation analysis.
To test whether these deviations could originate from our treatment of the halo bias, we repeat our measurements taking the fourth and the first mass bin for the MS and the MXXL, respectively, as our density tracers for all samples in each simulation. We re-calculate $w_\mathrm{g+}^*$ as the cross-correlation between the new density tracer and the shape sample in the respective mass bins, and derive $r_\mathrm{g+}$ using $w_\mathrm{gg}^*$ of the new density tracer. We present the results in Fig.~\ref{fig:ratio}. We do not observe any significant changes to the trends at the low- and high-mass ends and hence conclude that our removal of halo bias is robust. The deviation from our theoretical prediction at low masses appears even slightly more significant. It also suggests that the integral constraint is accounted for correctly, and that the deviations are driven by the alignment rather than the clustering of haloes.

\begin{table*}
	\centering
	\caption{Mean and 68\% confidence interval of the power-law fit parameters of the IA model from the likelihood analysis over the Millennium simulation, the Millennium-XXL simulation, their joint contribution, the Milennium simulation at $z = 0.46$, the Millennium simulation using the reduced inertia tensor (\textit{rit}) and real data. We exclude the highest-mass point in the MXXL only and joint fits. Note that the values from the snapshot at different redshift and from the reduced inertia tensor assumption are compatible with the outcomes of the MS only. A discussion about the reasons why the reduced $\chi^2$ values obtained considering the Millennium-XXL simulation (and, consequently, the joint analysis) significantly differ from unity is presented in the text.}
	\label{tab:param}
	\begin{tabular}{c||ccccc} % six columns, alignment for each
		\hline \hline
		\ & MS only & MXXL only & Joint & MS, $z=0.46$ & MS, \textit{rit} \\
		\hline
		$\beta_{\mathrm{M}}$					  & $0.31^{+0.03}_{-0.03}$   & $0.38^{+0.01}_{-0.01}$  & $0.36^{+0.01}_{-0.01}$ &  $0.35^{+0.03}_{-0.03}$ & $0.29^{+0.02}_{-0.02}$ \\
		$\log_{10} (A \ [\mathrm{Mpc}/h]^{1/2})$ & $-0.96^{+0.04}_{-0.05}$ & $-0.947^{+0.003}_{-0.003}$ & $-0.942^{+0.003}_{-0.003}$&$-0.72^{+0.04}_{-0.04}$&  $-1.21^{+0.04}_{-0.04}$\\
		$\chi^2 / \mathrm{d.o.f.}$			  & $0.27$                                 & $7.71$			 & $4.80$			    & $2.27$			      & $0.70$	 \\
		\hline \hline
		\ & Real data & \ & \ \\
		\hline
		$\beta_{\mathrm{M}}$ & $0.56^{+0.05}_{-0.05}$ & &\\
		$\log_{10} A_r $ & $0.61^{+0.03}_{-0.04}$ & & & & \\
		$\chi^2 / \mathrm{d.o.f.}$			  & $1.68$  & & & & \\
		\hline \hline 
		\end{tabular}
		
\end{table*}

The decreased alignment strength in the $10^{14.5}-10^{15} M_\odot$ bin could be caused by a substantial fraction of haloes still in the process of formation at $z=0$, violating the assumption of virial equilibrium in our model. 
At the low-mass end, instead, alignments are stronger than predicted by the model, which suggests that a different, or additional, alignment mechanism is at play. While our MS halo sample does not contain satellites, defined here as gravitationally bound sub-haloes, many haloes of Milky Way mass and less could be infalling along filaments onto larger haloes, a process that is distinct from our modelling ansatz (\citealt{ForeroRomeroetal2014}, and references therein). We assumed that the alignment is a perturbative effect to a largely randomly oriented intrinsic ellipticity $\epsilon$ (see Eq.~\ref{eq:epsilon}), which is unlikely to hold in an infall/filament alignment scenario. Indeed, we see evidence that the intrinsic halo ellipticity affects the mass scaling of the alignment signal, violating a tenet of our model. We demonstrate this by normalising the components of the ellipticity of each halo to 
\begin{align}
\label{eq:epsilonprime}
    	\epsilon_{1}' = \frac{\epsilon_1}{\sqrt{\epsilon_1^2 + \epsilon_2^2}} = \cos{2\phi}\ , \\ 
\label{eq:epsilonprimesecond}
           \epsilon_{2}' = \frac{\epsilon_2}{\sqrt{\epsilon_1^2 + \epsilon_2^2}} = \sin{2\phi}\ ,
\end{align}
where $\phi$ is the polar angle of the major axis of the ellipse in the Cartesian coordinate system of the simulation box. Near-spherical haloes with $|\epsilon|<0.01$ are excluded from further analysis. We then re-measure $w_\mathrm{g+}^*$ and $r_\mathrm{g+}$, which are multiplied by the mean ellipticity of the haloes in each bin, to make the signals directly compatible with the original results. As can be seen in Fig.~\ref{fig:ratio}, the mean-ellipticity $r_\mathrm{g+}$ is generally about 15\% below the original signals, which implies that haloes with above-average ellipticity drive the alignment. Significant deviations arise for the second and third mass bin of the MS, confirming a breakdown of our model on the scales of galaxies.

\subsection{Observation data}
\label{subsec:obsdatard}
\begin{figure}
	\centerline{	
	\includegraphics[scale = 0.7, keepaspectratio]{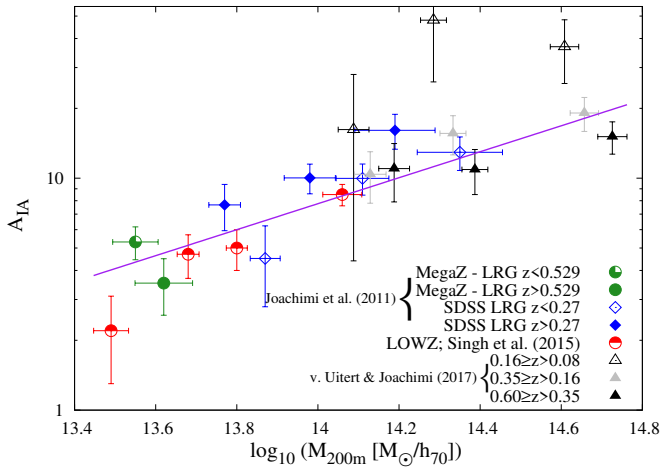}}
	\caption{Amplitude of the intrinsic alignment model as a function of halo mass from observations. The results from \citet{Joachimietal2011} and \citet{Singhetal2015} are for luminous red galaxies, while \citet{vanUitertJoachimi2017} measured the signal for galaxy clusters. The solid purple line shows the best-fit line from our likelihood analysis. The exact values of the parameters are shown in Table~\ref{tab:param}.}
	\label{fig:realdata}
\end{figure}

We repeat the likelihood analysis for the collection of observational datasets shown in \citet[figure 7]{vanUitertJoachimi2017}: we consider all 21 data points, which include literature results for LOWZ LRGs from \citet{Singhetal2015}, MegaZ-LRGs and SDSS LRGs from \citet{Joachimietal2011}, as well as results for the clusters contained in the redMaPPer catalogue version 6.3 from \citeauthor{vanUitertJoachimi2017}. We neglect the error bars on the mass, which are smaller than the errors on $A_{\mathrm{IA}}$ and whose impact is subdominant (\citeauthor{vanUitertJoachimi2017}), and treat all the data as independent, as in \citeauthor{vanUitertJoachimi2017}.
% the noise impact is hard to model, since the amplitude errors have all been estimated with different approaches. 
In this way, the covariance matrix is diagonal. We show the points, together with the best-fitting power-law from our analysis, in Fig.~\ref{fig:realdata}.

We adopt the model of Eq.~(\ref{eq:modelrg+}) but with a different prefactor $A_r$, which has an altered meaning and is now dimensionless, thus making it impossible to directly compare its value to the one obtained with simulation data. We also assume a different range for the flat prior in the fit for this new parameter, namely $\log_{10} A_r \in [0.4;0.9]$. 
The outcomes of our analysis are shown in Fig.~\ref{fig:post}\protect\subref{fig:postreal} and in Table~\ref{tab:param}: we observe that the value of the reduced chi-square for this latter analysis can be improved to 1.36 by excluding the high-redshift SDSS results (filled-blue diamonds) without affecting the value of the slope in a significant way.

The incompatibility between the values of the slope $\beta_{\mathrm{M}}$ between simulation and real data is significant. This discrepancy is likely to be attributed to an additional mass dependence of the response of a galaxy ellipticity to the ellipticity of its host halo or the local tidal field, and, more generally, to the fact that, while we observe luminous matter, the simulations and the theory model only consider dark matter. Moreover, the effective scale at which ellipticities are measured in observational data changes from the central part of galaxies (since weak lensing techniques are employed) at the low-mass end, to the distribution of satellites in clusters that trace the halo out to the virial radius. We also note that a bias in $\beta_\mathrm{M}$ could be determined from data if, with increasing halo mass, there is also a trend in redshift, which however is not the case of the data collection of Fig.~\ref{fig:realdata}. 
%A more detailed discussion about the reasons that could explain this disagreement is presented in the next section.

\section{Conclusions}
In this work we studied the dependence of the intrinsic alignment amplitude on the mass of dark matter haloes, using data from the Millennium and Millennium-XXL $N$-body simulations.
 We derived the intrinsic alignment amplitude scaling with mass in the tidal alignment paradigm for a dark matter-only universe. Our analytical estimate assumed a virialised system which is weakly perturbed by ambient tidal shear fields. In this model, the magnitude of the tidal distortion of a halo is determined entirely by the amount of fluctuations of the tidal fields on the mass scale of the halo. The model predicts a scaling with halo mass $\propto M^{\beta_{\mathrm{M}}}$, with $\beta_{\mathrm{M}} \in [1/3, 1/2]$, if we approximate the matter power spectrum with a power-law. 

We mimicked the observational approach to measure the halo shape-position alignments, and we performed a Bayesian analysis on the mass dependence of the alignments with a simple power-law model. We found that the results from the two simulation data sets agree very well in the mass range covered by both. Our model predicts the overall trend of the mass scaling of halo alignments remarkably well, but we observed significant deviations from the predictions at masses below $\sim 10^{13} M_\odot$ as well as in the highest mass bin near $10^{15} M_\odot$. We demonstrated that those are not caused by our analysis choices, nor by changes of the slope of the matter power spectrum as a function of scale. Rather, we found evidence that the physical processes driving the halo alignments in these mass ranges violate the assumptions of our simple model. 

The joint analysis of the Millennium and Millennium-XXL simulations yields $\beta_{\mathrm{M}} = 0.36^{+0.01}_{-0.01}$, and for cluster-size scales, for which we predicted $\beta_{\mathrm{M}} = 0.36$, we found $\beta_{\mathrm{M}} =0.38^{+0.01}_{-0.01}$, however generally with bad goodness of fit due to the aforementioned deviations. Furthermore, there is no obvious dependence of the mass scaling on redshift or on the definition of the inertia tensor which describes the shape of the halo.

We repeated our statistical analysis using observational data, inferring a value of $\beta_{\mathrm{M}} = 0.56^{+0.05}_{-0.05}$, which is not compatible with the simulation results or our model prediction. We argue that the incompatibility is attributed to the fact that simulations consider a dark-matter only universe, while we observe luminous matter: hydrodynamical simulations suggest in fact that the presence of baryons could significantly affect the shape of the host halo, especially in the inner regions (\citealt{Kiesslingetal2015}, and references therein). 

For example, \citet{Bailinetal2005} found that baryons cause haloes to be closer to spherical but that their impact diminishes at larger radii (and thus with larger mass values), in agreement with \citet{Kazantzidisetal2004}, and more recently \citet{Tennetietal2014} and \citet{Velliscigetal2015a} discovered that galaxies are more misaligned with the hosting halo at lower masses. Also, \citet{Tennetietal2015} compared the projected density-shape correlation function in three mass bins covering four orders of magnitude between a hydrodynamical and a dark matter-only simulation, showing that for the lowest mass bin the stellar matter signal is more than halved with respect to the dark matter signal, while at higher mass the fractional difference between the two does not exceed $40\%$.Finally, the effective scale at which observations evaluate ellipticities moves outwards with increasing mass. All these trends contribute to a steepening of the mass scaling compared to dark matter halo alignments.

More quantitatively, \citet{Okumuraetal2009} showed the presence of a significant amount of misalignment between the orientations of LRGs and their dark matter haloes. They estimated a typical misalignment angle of $\sigma_\theta= 34.9°^{+1.9}_{-2.1}$ degrees, which causes the galaxy-shape correlation function to be about half the halo-shape correlation function on all scales. We used this result to rescale the MXXL $r_{\mathrm{g+}}$ values according to their mass, retrieving the median cosine of the misalignment angle for the highest mass bins from \citet[figure 8]{Velliscigetal2015a}. With this approach, we obtain $\beta_\mathrm{M} = 0.66^{+0.01}_{-0.01}$, a significantly higher value with respect to what we found using $N$-body simulations, and compatible within $2\sigma$ with the observational outcome.

These results hint that this discrepancy could be addressed by measuring the slope $\beta_{\mathrm{M}}$ considering a hydrodynamical simulation large enough to contain clusters, which could then account for the additional effects of baryons and gas. 

After completing this analysis, another study of the mass scaling of intrinsic alignments in $N$-body simulations appeared. \citet{Xiaetal2017} also develop an analytic model based on the tidal alignment paradigm, incorporating a mass dependence solely via the halo bias that enters by employing the halo power spectrum in lieu of the matter power spectrum. The authors find good agreement between predicted and measured halo ellipticity-ellipticity correlation functions. It will be interesting to compare the two ans\"{a}tze in future work.

\section*{Acknowledgements}
We thank Raul Angulo for helpful information about the Millennium-XXL catalogue. DP acknowledges support by an Erasmus+ traineeship grant. BJ acknowledges support by an STFC Ernest Rutherford Fellowship, grant reference ST/J004421/1. MB acknowledges the support of the OCEVU Labex (ANR-11-LABX-0060) and the A*MIDEX project (ANR-11-IDEX-0001-02) funded by the ``Investissements d'Avenir'' French government program managed by the ANR. SH acknowledges support by the DFG cluster of excellence \lq{}Origin and Structure of the Universe\rq{} (\href{http://www.universe-cluster.de}{\texttt{www.universe-cluster.de}}). EvU acknowledges support from an STFC Ernest Rutherford Research Grant, grant reference ST/L00285X/1. This work was granted access to the HPC resources of Aix-Marseille Universit\'e financed by the project Equip@Meso (ANR-10-EQPX-29-01) of the program ''Investissements d'Avenir`` supervised by the Agence Nationale pour la Recherche (ANR). 

%%%%%%%%%%%%%%%%%%%%%%%%%%%%%%%%%%%%%%%%%%%%%%%%%%

%%%%%%%%%%%%%%%%%%%% REFERENCES %%%%%%%%%%%%%%%%%%

% The best way to enter references is to use BibTeX:

\bibliographystyle{mnras}
\bibliography{biblio} % if your bibtex file is called example.bib

\bsp	% typesetting comment
\label{lastpage}
\end{document}